# A Matroid Framework for Noncoherent Random Network Communications

Maximilien Gadouleau, *Member, IEEE* and Alban Goupil

## Abstract

Models for noncoherent error control in random linear network coding (RLNC) and store and forward (SAF) have been recently proposed. In this paper, we model different types of random network communications as the transmission of flats of matroids. This novel framework encompasses RLNC and SAF and allows us to introduce a novel protocol, referred to as random affine network coding (RANC), based on affine combinations of packets. Although the models previously proposed for RLNC and SAF only consider error control, using our framework, we first evaluate and compare the performance of different network protocols in the error-free case. We define and determine the rate, average delay, and throughput of such protocols, and we also investigate the possibilities of partial decoding before the entire message is received. We thus show that RANC outperforms RLNC in terms of data rate and throughput thanks to a more efficient encoding of messages into packets. Second, we model the possible alterations of a message by the network as an operator channel, which generalizes the channels proposed for RLNC and SAF. Error control is thus reduced to a coding-theoretic problem on flats of a matroid, where two distinct metrics can be used for error correction. We study the maximum cardinality of codes on flats in general, and codes for error correction in RANC in particular. We finally design a class of nearly optimal codes for RANC based on rank metric codes for which we propose a low-complexity decoding algorithm. The gain of RANC over RLNC is thus preserved with no additional cost in terms of complexity.

## I. Introduction

During transmission through a network, the data can be modified, as in network coding, without affecting their decoding. However, other modifications, such as packets in error or lost, corrupt the nature

M. Gadouleau is with the Department of Computer Science, Queen Mary, University of London, London E1 4NS, UK and on leave from CReSTIC, Université de Reims Champagne-Ardenne, Reims, France. A. Goupil is with CReSTIC, Université de Reims Champagne-Ardenne, Reims, France. E-mails: mgadouleau@eecs.qmul.ac.uk, alban.goupil@univ-reims.fr

 

of the transmitted message. Recently, operator channels have been proposed to differentiate these two types of modifications for data transmission using random linear network coding (RLNC) [1] and store and forward (SAF) [2], respectively. For RLNC, it is shown that data transmission is equivalent to the communication of a linear subspace of a given vector space [1]; for SAF, however, a subset of a set is transmitted [2]. Using these operator channels, noncoherent error correction in RLNC and SAF can be reduced to coding theoretic problems on linear subspaces and subsets, respectively.

In this paper, we generalize the models described above by viewing random data transmission through a network as the communication of a flat of a matroid. Matroids [3] can be viewed as the combinatorial essence of independence, and hence are a generalization of linear independence; flats of a matroid can be viewed as generalizations of linear subspaces. Studying matroids allows to focus on the combinatorial aspects of independence and combinations, without assuming any underlying algebraic structure. Although the models for RLNC and SAF were introduced for error control, our matroid framework allows us to study protocols for both the error-free case and the case where error control is considered. The matroids associated to RLNC and SAF are easily determined and are well-known. In particular, we shall show that the matroid for RLNC—the projective geometry—only considers a fraction of packets (around $q^{-1}$, where packets are viewed as vectors over $\mathrm{GF}(q)$), hence leading to a rate loss of around one symbol per packet.

In order to thwart this rate loss, we introduce a new way to combine packets for network coding, referred to as random affine network coding (RANC), where packets are viewed as points instead of vectors and new packets are created via affine combinations. The associated matroid is the well-known and thoroughly studied affine geometry, whose flats are affine subspaces of an affine space. Unlike RLNC, which only considers a fraction of all packets, RANC works on all possible packets, thus utilizing a better encoding of messages into flats. Moreover, since affine combinations are particular linear combinations, the complexity at the intermediate nodes is not increased. At the receiver end, the message can be decoded using Gaussian elimination, for an affine subspace is no more than a translated linear subspace. Therefore, utilizing RANC instead of RLNC does not increase the complexity at the source, the intermediate nodes, or the destinations.

Then, using our matroid framework, we determine, evaluate, and compare the performances of different network protocols. We first define the data rate of a matroid as the ratio between the amount of information carried by the flat, i.e. the logarithm of the number of flats, and the size of the message transmitted through the network. We also investigate the average delay of a matroid, which reduces to the coupon collector problem for SAF. Combining these two parameters, we also define the throughput of a matroid as the proportion of useful information received by the destination. We show that RANC outperforms RLNC in terms of data rate, while offering a similar average delay, thus yielding a higher throughput of around one symbol per packet. We then study the delay in more detail via the average number of independent packets in a given number of received packets. We hence demonstrate that this number tends to the optimum for RLNC and RANC when the field size increases, while the number of independent packets in SAF follows an exponential recovery. We finally investigate the possibilities of partial decoding. We prove that partial decoding is highly unlikely in RLNC and RANC, while for SAF all packets are decodable. Therefore, RLNC and RANC follow a zero-one pattern: no packets can be decoded before receiving the total number of packets in a message, and once this amount is received, the whole message can be decoded. On the other hand, SAF follows an exponential recovery in terms of partially decodable packets.

The study described above considers an error-free transmission through the network. In the presence of errors, message alterations (packets lost, injected, or in error) correspond to modifications of the transmitted flat. The network can hence be viewed as an operator channel, which generalizes the channels defined in [1] and [2] for RLNC and SAF, respectively. We then introduce two metrics for error correction in random network communications. These metrics, referred to as the lattice distance and the modified lattice distance, respectively, are identified with previously proposed metrics for RLNC [4] and SAF [2]. We also place constant-dimension codes used for RLNC [1] and constant-weight codes used for SAF [2] into the new framework of matroid codes, which are codes on flats of a matroid sharing the same rank. We then investigate error control for RANC with codes on affine subspaces. We derive bounds on the maximum cardinality of such codes and determine a nearly optimal class of codes based on liftings of rank metric codes [5]. We finally design a decoding algorithm for these codes based on, and with the same order of complexity as the decoder proposed in [1] for RLNC. The rate gain of RANC over RLNC is therefore preserved when error correction is considered, and at no additional cost in terms of complexity.

We summarize the advantages of our matroid framework below.

- First, this framework is very general, and offers a unified approach for distinct problems such as SAF, RLNC, and RANC. It offers to focus on the combinatorial properties of network protocols, in terms of both combinations and encoding. Also, associating a matroid to a protocol provides with a new tool to study and compare the performances of different protocols for both the error-free case and when error control is enforced.

- Second, different properties of a protocol arising from matroid theory can be discovered. For example, we demonstrate how RLNC can be viewed as a matroid on only a fraction of all possible packets, and hence determine the data rate and the actual number of possible combinations offered by RLNC. The lattice distance also illustrates the easiest way to alter a message, hence highlighting the sensitivity of network coding to errors.

- Third, when studying error control, the advantages of using an operator channel still apply to our general framework. Although the matroid depends on the protocol, it is independent of the actual network, rendering our approach noncoherent and robust to network topology changes. Moreover, errors on the message level, such as packets lost or injected, and errors on the packet level (bits or symbol errors) can be detected and corrected using the same class of codes. The problem of error control can be eventually tackled using methods from algebraic coding, such as binary constant-weight codes or rank metric codes.

- Fourth, our model offers a wealth of alternatives to the protocols already proposed in the literature, as many different types of matroids have been previously discovered and studied. One of these alternatives introduced here, RANC, is shown to outperform RLNC. Different matroids may lead to different tradeoffs between the number of possible combinations and the data rate. Also, it is known that linear network coding is not optimal when multiple sources are considered [6], then non-linearly representable matroids may offer a higher throughput than RLNC or RANC in these cases.

The rest of the paper is organized as follows. Section II reviews some necessary backgrounds on matroids and error correction models for RLNC and SAF. In Section III, we introduce the model based on matroids for error-free communications. Section IV introduces and illustrates random affine network coding. In Section V, we evaluate and compare the different performance parameters of matroids. In Section VI, we model the alterations of the message into an operator channel, and study the codes used for error correction in random network communications. Finally, Section VII details possible extensions of our work.

## II. Preliminaries

### A. Matroids

We review below the definition and major properties of matroids and their flats. Although the concepts introduced below arise from matroid theory, they all are generalizations of well-known concepts in linear algebra. For an extensive account on matroid theory, the interested reader is referred to [3].

For any set $E$, we denote the set of subsets of $E$ with cardinality $0 \leq i \leq |E|$ as $\mathcal{P}(E, i)$ and its power set as $\mathcal{P}(E) = \bigcup_{i=0}^{|E|} \mathcal{P}(E, i)$. A *matroid* is a pair $\mathcal{M} = (E, \mathcal{I})$, where $E$ and $\mathcal{I} \subseteq \mathcal{P}(E)$ are referred to as the *ground set* and the *independent sets* of $\mathcal{M}$, respectively. The independent sets are generalizations of linearly independent vectors and satisfy the following three axioms: $\varnothing \in \mathcal{I}$; if $A \in \mathcal{I}$ and $B \subset A$, then $B \in \mathcal{I}$; if $I_1, I_2 \in \mathcal{I}$ with $|I_1| > |I_2|$, then there exists $e \in I_1 \backslash I_2$ such that $I_2 \cup \{e\} \in \mathcal{I}$. The third axiom, referred to as the *independence augmentation axiom*, is crucial as it guarantees that any family of independent elements can be extended to form a *basis* (a maximal family of independent elements). Clearly, all bases have the same cardinality.

To any matroid is associated a *rank* function $\mathrm{rk}(A)$ for all $A \subseteq E$, defined as the maximum number of independent elements in $A$. For any two subsets $A, B \subseteq E$, we have the *submodular inequality* $\mathrm{rk}(A \cup B) + \mathrm{rk}(A \cap B) \leq \mathrm{rk}(A) + \mathrm{rk}(B)$. The rank of a matroid is simply the rank of its ground set, and is the number of elements in any basis. The *closure* $\mathrm{cl}(A)$ of a subset $A$ of the ground set is then defined as the maximal subset $B \subseteq E$ such that $B$ contains $A$ and $\mathrm{rk}(B) = \mathrm{rk}(A)$. The closure is unique, as it can be shown that $\mathrm{cl}(A) = \{e \in E : \mathrm{rk}(A \cup e) = \mathrm{rk}(A)\}$ [3, Eq. 1.4.1]. The closure and the rank are generalizations of the span of a set of vectors and the dimension of that span, respectively.

A *flat* is a set equal to its closure, which is a generalization of a linear subspace. In particular, we refer to any flat of rank $r - 1$ in a matroid of rank $r$ as a *hyperplane*. By extension, we refer to any family of $k$ independent elements in a flat of rank $k$ as a basis of that flat. The set of flats of a matroid, denoted as $\mathcal{F}$, is closed under intersection. We also denote the set of flats of rank $k$ as $\mathcal{F}_k$ and its cardinality as $N_k$ for any $0 \leq k \leq r$. Furthermore, the set of flats ordered by inclusion forms a lattice (in the partially ordered set sense), where the meet of two flats is their intersection, and their join is the closure of their union.

A matroid may contain loops and parallel elements. A *loop* $l$ is an element of the ground set belonging to no independent set: $\{l\} \notin \mathcal{I}$; alternatively, $l$ belongs to the closure of the empty set. A collection of elements are said to be *parallel* if they are pairwise dependent: $\{e_i, e_j\} \notin \mathcal{I}$ for $i \neq j$; they hence all belong to a set with rank 1. A loop and parallel elements are generalizations of the all-zero vector and collinear vectors, respectively. A matroid is said to be *simple* if it does not contain any loops or parallel elements. For any matroid $\mathcal{M}$, the simple matroid obtained by removing all loops and keeping only one element in each set of parallel elements of $\mathcal{M}$ has the same lattice of flats as $\mathcal{M}$. For any simple matroid, we have $\mathcal{F}_0 = \{\emptyset\}$, $N_0 = 1$ and $\mathcal{F}_1 = \mathcal{P}(E, 1)$, $N_1 = |E|$.

We now review three important classes of matroids. First, the *free matroid* on $r$ elements, classically denoted as $U_{r,r}$, has $[r] = \{0, 1, \ldots, r-1\}$ as a ground set, and any subset of $[r]$ is independent. Clearly, this matroid is simple, has rank $r$, and any subset of $[r]$ is a flat.

Second, the *projective geometry* $PG(r - 1, q)$ has all the non-zero vectors of $\mathrm{GF}(q)^r$ with leading nonzero coefficient equal to 1 as ground set, where linear independence is used. This matroid is simple, has rank $r$ and its flats are in one-to-one correspondence with the linear subspaces of $\mathrm{GF}(q)^r$. Therefore, the number of flats of rank $k$ is given by the Gaussian binomial $\begin{bmatrix} r \\ k \end{bmatrix} = \prod_{i=0}^{k-1} \frac{q^{r-i}-1}{q^{k-i}-1}$, which satisfies $q^{k(r-k)} \leq \begin{bmatrix} r \\ k \end{bmatrix} < K_q^{-1} q^{k(r-k)}$ for all $0 \leq k \leq r$, where $K_q = \prod_{i=1}^{\infty}(1-q^{-i}) < 1$ tends to 1 when $q$ tends to infinity [7].

Third, removing a hyperplane from $PG(r-1,q)$ yields the *affine geometry* $AG(r-1,q)$. This matroid is also simple with rank $r$ and its flats are the affine subspaces of $\mathrm{GF}(q)^{r-1}$; there are $q^{r-k}\begin{bmatrix} r-1 \\ k-1 \end{bmatrix}$ flats of rank $k$ for all $0 \leq k \leq r$. Any affine subspace with rank $k$ can be represented by a linear subspace with dimension $k-1$ translated by a point belonging to a complementary linear subspace. By definition, $AG(r-1,q)$ is a submatroid of $PG(r-1,q)$, and can be viewed as a matroid on the points in $\mathrm{GF}(q)^{r-1}$, where two points $\mathbf{u}, \mathbf{v}$ are affinely independent if and only if the vectors $(1,\mathbf{u}), (1,\mathbf{v}) \in \mathrm{GF}(q)^r$ are linearly independent.

### B. Error control for RLNC and SAF

We now review the existing models for error correction in RLNC and SAF given in [1] and [2], respectively. For RLNC, several techniques have been proposed for error correction (see [8], [9] for coherent error correction); however, we are interested here in the operator channel approach introduced in [1] for noncoherent error control. Suppose a message, encoded into $k$ linearly independent packets in $\mathrm{GF}(q)^n$, is transmitted through a network using RLNC. Since the linear combinations operated by the intermediate nodes do not modify the subspace spanned by the packets, RLNC is viewed as the transmission of a linear subspace of dimension $k$ of $\mathrm{GF}(q)^n$. The alterations of the message (packets lost, injected, or in error) hence correspond to modifications of that subspace. The transmission of a message using RLNC is hence modeled as an operator channel which modifies the input subspace sent by the source into the output subspace received by the destination. Accordingly, codes on subspaces, and more especially codes on a Grassmannian referred to as *constant-dimension codes*, have been proposed for error correction in RLNC. Two metrics between subspaces have been proposed: the subspace metric and the injection metric [4]. The maximum cardinality of a constant-dimension code, consisting of subspaces of $\mathrm{GF}(q)^n$ with dimension $k$, with minimum injection distance $d$ (and equivalently, minimum subspace distance $2d$) is between $q^{\min\{k(n-k-d+1),(n-k)(k-d+1)\}}$ and $K_q^{-1} q^{\min\{k(n-k-d+1),(n-k)(k-d+1)\}}$. These bounds follow the Singleton bound in [1] and the inequalities on the Gaussian binomial above and were tightened in [10]–[12].

A possible and highly practical construction of constant-dimension codes, referred to as liftings of rank metric codes, has been proposed in [13]. Rank metric codes [5], [14], [15] are codes on matrices in $\mathrm{GF}(q)^{k \times \nu}$, where the rank distance between two matrices is simply the rank of their difference. The number of matrices with rank $r$ in $\mathrm{GF}(q)^{k \times \nu}$ is given by $\genfrac{[}{]}{0pt}{}{k}{r} \prod_{i=0}^{r-1}(q^\nu - q^i)$ [5]. The maximum cardinality of a rank metric code in $\mathrm{GF}(q)^{k \times \nu}$ with minimum rank distance $d$ is given by $q^{\min\{k(\nu-d+1), \nu(k-d+1)\}}$ and is achieved by *Gabidulin codes* [5], an analogue of Reed-Solomon codes. For any $\mathbf{M} \in \mathrm{GF}(q)^{k \times \nu}$, the linear lifting $I_\mathcal{L}(\mathbf{M})$ of $\mathbf{M}$ is the row space of the matrix $(\mathbf{I}_k | \mathbf{M})$, a subspace of $\mathrm{GF}(q)^{k+\nu}$ with dimension $k$ [13]. The injection distance between two liftings of matrices is equal to the rank distance between the matrices, hence the lifting of a rank metric code has the same minimum injection distance as the original code. In particular, lifings of Gabidulin codes are nearly optimal constant-dimension codes for which low-complexity decoding algorithms were proposed [1], [13].

Similarly, an operator channel has been proposed for error correction in SAF in [2]. Suppose $k$ packets in $[q]^n$ are transmitted through a network with SAF, where we denote $[q] = \{0, 1, \ldots, q-1\}$ for any integer $q$. Also, assume the packets arrive at the destination in a different order to which they were sent in. Then only the set of packets is preserved, and SAF is modeled as the transmission of a subset of cardinality $k$ of $[q^n]$. Codes on subsets have hence been proposed for error control in SAF with two distinct metrics: the Hamming metric and the modified Hamming metric. Since subsets of $[q^n]$ are in bijection with vectors in $\mathrm{GF}(2)^{q^n}$, codes on subsets can be viewed as binary codes; in particular, codes on subsets with the same cardinality can be viewed as binary constant-weight codes.

Similarly to the case of constant-dimension codes, a practical construction of constant-weight codes with length $q^n$ and weight $q^l$ is the lifting of a nonrestricted Hamming metric code in $\mathrm{GF}(q^{n-l})^{q^l}$. The lifting $I_\mathcal{S}(\mathbf{X})$ of any word $\mathbf{X} = (X_0, X_1, \ldots, X_{q^l-1}) \in [q^{n-l}]^{q^l}$ is obtained by added the header $i$, encoded into $l$ symbols of $[q]$, in front of the packet corresponding to $X_i$ for all $0 \le i \le q^l - 1$. Alternatively, it is the subset $\{x_0, x_1, \ldots, x_{q^l-1}\} \in \mathcal{P}([q^n], q^l)$, where $x_i = iq^{n-l} + X_i$ for $0 \le i \le q^l - 1$. The lifting $I_\mathcal{S}$ preserves the Hamming distance: $d_\mathrm{H}(I_\mathcal{S}(\mathbf{X}), I_\mathcal{S}(\mathbf{Y})) = 2d_\mathrm{H}(\mathbf{X}, \mathbf{Y})$, and liftings of nonrestricted Hamming metric codes can be used for error control with SAF.

## III. Transmission model

### A. Model and discussion

In this section, we introduce a noncoherent communication model based on matroids for error-free data transmission through a network. We consider a source wishing to transmit a message $M$ in the alphabet $[A] = \{0, 1, \ldots, A-1\}$ through a network toward a set of destinations. Let $(E, \mathcal{I})$ be a simple matroid and denote its set of flats of rank $k$ as $\mathcal{F}_k$ for all $k$, and assume that both the source and the destination know a common injective map $G$ from $[A]$ to $\mathcal{F}_k$.

The error-free data transmission follows three steps.

- Step I: at the source. The source encodes the original message $M$ into a flat $f = G(M) \in \mathcal{F}_k$. Then a stream of elements of $f$ containing a basis of $f$ is transmitted into the network.

- Step II: in the network. Each intermediate node combines the elements it has previously received by selecting and retransmitting elements of their closure.

- Step III: at each destination. The destination waits until it receives a basis of $f$, and then recovers the original message by determining $M = G^{-1}(f)$.

We now provide several remarks regarding the matroids and the flats used in our model. First, we consider flats of a matroid, for the matroid structure ensures that the rank function is well-behaved. Indeed, a flat of rank $k$ can only be described by $k$ independent elements, no less and no more. Also, the independence augmentation axiom reviewed in Section II-A guarantees that any set of less than $k$ independent elements can be extended into a basis of $k$ elements of the flat.

Second, a non-simple matroid contains loops and parallel elements. By definition, a loop belongs to every flat and hence does not carry any information about the transmitted flat. Also, two parallel elements belong to the same flats and are combined in the same way, hence they carry the same information. Therefore, loops and parallel elements are unnecessary to the destination, and our assumption of considering simple matroids only does not lead to any loss of generality.

Third, although flats of any rank may be sent, the following two reasons justify our assumption to send flats of the same rank $k$ only. Foremost, no transmitted flat is properly contained by another, thus rendering the decoding non-ambiguous. Also, the destination always expects the same number of independent elements to start decoding, hence simplifying the decoding process.

Fourth, the number of possible combinations for an intermediate node is given by the cardinality of the closure of the elements it has received. However, not all flats of the same rank necessarily have the same cardinality and the same number of bases, which results in different protections to packet losses. However, as shown below, SAF and RLNC use matroids for which all flats of the same rank have equal cardinalities. Matroids satisfying this property are referred to as *perfect matroid designs* [16], [17, Section 3.4]. Due to their highly specific structure, very few classes of perfect matroid designs are known so far. When considering a perfect matroid design, we shall denote the cardinality of any flat of rank $k$ as $C_k$ henceforth, where $C_0 = 0$ and $C_1 = 1$ for any simple perfect matroid design.

We also comment on the validity of our model and on some practical issues regarding its realization. Our model is general and does not take advantage of any knowledge of the network topology. It is hence *noncoherent*, and is robust to network topology variations, such as node or link appearance/disappearance.

Accordingly, the intermediate nodes are assumed to operate blindly on the elements they receive, regardless of the source, the destination, or the actual transmitted data.

In terms of practical implementation, without loss of generality, we assume that an element is encoded in one packet of length $n$ over $\mathrm{GF}(q)$. All intermediate nodes should have an efficient algorithm to combine elements; this combination algorithm, however, does not guarantee to yield a new basis of the flat. This operation can be viewed as a form of random sampling on the elements of a flat. Also, the destination needs an efficient algorithm to retrieve the original message from any basis of the flat. For a general matroid, efficient algorithms may not exist; however, we shall only consider matroids for which combining elements can be done efficiently.

### B. Matroids for SAF and RLNC

We now determine the matroids associated to SAF and RLNC.

First, for SAF, the only combination possible is the selection of an element, hence the flats are the subsets of cardinality $k$ of $[q^n]$. The associated matroid is the free matroid $U_{q^n, q^n}$ and we have $E = [q^n]$, $\mathcal{F}_k = \mathcal{P}(E, k)$ for all $0 \leq k \leq q^n$, and hence $N_k = \binom{q^n}{k}$ and $C_k = k$. In order to use notations reflecting the protocol and the alphabet and length of packets, we denote $U_{q^n, q^n}$ as $\mathcal{S}(q, n)$ or simply $\mathcal{S}$ when there is no ambiguity.

Second, our model differs slightly from the purely random linear combinations typically proposed for RLNC. Indeed, a linear combination may yield the all-zero vector or collinear vectors: these are respectively loops and parallel elements. However, our model considers the simple matroid associated to RLNC, which is the projective geometry $PG(n-1, q)$. Its ground set $E$ is the set of one-dimensional subspaces of $\mathrm{GF}(q)^n$, and $\mathcal{F}_k$ is a Grassmannian for $0 \leq k \leq n$, and hence $N_k = \begin{bmatrix} n \\ k \end{bmatrix}$ and $C_k = \begin{bmatrix} k \\ 1 \end{bmatrix}$. Clearly, the combinations operated by the intermediate nodes are linear combinations which ensure the output vector is non-zero and has leading non-zero coefficient equal to 1, while decoding the message at the destination is achieved via Gaussian elimination. We denote $PG(n-1, q)$ as $\mathcal{L}(q, n)$ or simply $\mathcal{L}$ when there is no ambiguity henceforth.

## IV. Random affine network coding

In this section, we introduce a novel network coding scheme, referred to as *random affine network coding* (RANC), where packets are viewed as points in an affine space and where intermediate nodes combine packets by affine combinations. An affine combination of points $\mathbf{v}_0, \mathbf{v}_1, \ldots, \mathbf{v}_{k-1} \in \mathrm{GF}(q)^n$ is any sum of the form $\sum_{i=0}^{k-1} a_i \mathbf{v}_i$, where the scalars $a_i \in \mathrm{GF}(q)^n$ satisfy $\sum_{i=0}^{k-1} a_i = 1$. In other words, an affine combination corresponds to determining the centroid of the points $\mathbf{v}_i$ with masses $a_i$.

The set of all possible affine combinations of a collection of points, referred to as the *affine hull*, forms an affine subspace. The matroid associated to RANC is hence given by the affine geometry $AG(n, q)$, which we will denote as $\mathcal{A}(q, n)$ or simply $\mathcal{A}$ if there is no ambiguity about the parameter values. A collection of points $\mathbf{v}_0, \mathbf{v}_1, \ldots, \mathbf{v}_{k-1} \in \mathrm{GF}(q)^n$ are said to be *affinely independent* if $\sum_{i=0}^{k-1} b_i \mathbf{v}_i \neq 0$ for all $b_i$s not all zero and satisfying $\sum_{i=0}^{k-1} b_i = 0$. By definition, the rank of a set of points is given by the number of affinely independent points, and is equal to the rank of their affine hull. For any $\mathbf{v}_0, \mathbf{v}_1, \ldots, \mathbf{v}_{k-1} \in \mathrm{GF}(q)^n$, we then have $\mathrm{rk}(\mathbf{v}_0, \mathbf{v}_1, \ldots, \mathbf{v}_{k-1}) = \mathrm{rank}(\mathbf{1}|\mathbf{V})$, where $\mathbf{V} = (\mathbf{v}_0^T, \mathbf{v}_1^T, \ldots, \mathbf{v}_{k-1}^T)^T$ and $\mathrm{rank}$ denotes the number of linearly independent rows of a matrix. The set $\mathcal{F}_k$ of flats of rank $k$ being the set of affine subspaces of rank $k$, we have $N_k = q^{n-k+1} \begin{bmatrix} n \\ k-1 \end{bmatrix}$ and $C_k = q^{k-1}$ for $1 \leq k \leq n+1$ [3, Section 6.2].

We now provide guidelines for the implementation of RANC. First, encoding messages (viewed as the rows $\mathbf{m}_i$ of a matrix $\mathbf{M} \in \mathrm{GF}(q)^{k \times (n-k+1)}$) into affinely independent points can be simply done by adding the header $\mathbf{I}'_k = (\mathbf{0}|\mathbf{I}_{k-1})^T$ to obtain $(\mathbf{I}'_k|\mathbf{M}) \in \mathrm{GF}(q)^{k \times n}$. We shall refer to this encoding as the *affine lifting* of the matrix $\mathbf{M}$. Second, since affine combinations are particular linear combinations, the complexity of using RANC at the intermediate nodes is no higher than using RLNC. Third, we describe the decoding algorithm at the destination, thus showing that this does not increase complexity either. Suppose the destination receives $k$ affinely independent points $\mathbf{v}_0, \mathbf{v}_1, \ldots, \mathbf{v}_{k-1}$, then the first $k$ columns of $(\mathbf{1}|\mathbf{V}) \in \mathrm{GF}(q)^{k \times (n+1)}$ are linearly independent. Therefore, Gaussian elimination on $(\mathbf{1}|\mathbf{V})$ yields $(\mathbf{I}_k|\mathbf{M}')$, where $\mathbf{M}' = (\mathbf{m}_0^T, (\mathbf{m}_1 - \mathbf{m}_0)^T, \ldots, (\mathbf{m}_{k-1} - \mathbf{m}_0)^T)^T$. The decoding is finished by adding $\mathbf{m}_0$ to all the other rows of $\mathbf{M}'$. The complexity of the algorithm is hence dominated by the inversion of a matrix of order $k$, which is similar to the complexity for RLNC. We finally note that the Gaussian elimination could be modified in order to obtain the matrix $(\mathbf{1}|\mathbf{I}'_k|\mathbf{M})$ directly.

As seen in Section III-B, the simple matroid associated to RLNC is the projective geometry with rank $n$, whose alphabet only has $\begin{bmatrix} n \\ 1 \end{bmatrix} \sim q^{n-1}$ elements. This implies a loss in terms of data rate, as the elements are not optimally encoded into packets of length $n$. Similarly, any linear subspace has $\begin{bmatrix} k \\ 1 \end{bmatrix} \sim q^{k-1}$ elements, which compared to the $q^k$ possible linear combinations, leads to a decrease in the number of possible combinations. These issues are immediate consequences of the existence of a loop (the all-zero vector) and parallel elements (collinear vectors). Unlike RLNC, the matroid associated to RANC has rank $n+1$ and $q^n$ elements. By construction, $\mathcal{A}(q, n) = AG(n, q)$ is a submatroid of $\mathcal{L}(q, n+1) = PG(n, q)$. However, we shall demonstrate in the following that $\mathcal{A}(q, n)$ behaves closely to $\mathcal{L}(q, n+1)$, hence virtually allowing to work on packets of length $n+1$ instead of $n$.

We illustrate the difference between RLNC and RANC by using the butterfly network, depicted in Figure 1, where the source $S$ wants to transmit two messages $\mathbf{m}$ and $\mathbf{n}$ over $\mathrm{GF}(q)$ to the destinations $D_1$ and $D_2$. First, suppose RLNC is used. The source then encodes these messages into linearly independent vectors with their first non-zero coordinate equal to one by adding the following headers: $\mathbf{x} = (10|\mathbf{m})$ and $\mathbf{y} = (01|\mathbf{n})$. The only linear combination of one vector is simply the vector itself; all the linear combinations of $\mathbf{x}$ and $\mathbf{y}$ can be expressed as $\mathbf{x} + a\mathbf{y}$, where $a \in \mathrm{GF}(q)$. There are hence $q$ combinations possible, and $q-1$ lead to successful decoding at the destinations (if $a \neq 0$), leading to a success probability of $\frac{q-1}{q}$, which tends to 1 for large $q$. We remark that our model is consistent with the typical approach of RLNC, which allows combinations of the type $a\mathbf{x} + b\mathbf{y}$ instead.

Now suppose RANC is used. The source then encodes the messages into affinely independent points by adding the following headers: $\mathbf{u} = (0|\mathbf{m})$ and $\mathbf{v} = (1|\mathbf{n})$. Note that the header is one symbol long only, illustrating the gain of one symbol per packet of utilizing RANC over RLNC. The only affine combination of one point is the point itself; all the affine combinations of $\mathbf{u}$ and $\mathbf{v}$ can be expressed as $b\mathbf{u} + (1-b)\mathbf{v}$, where $b \in \mathrm{GF}(q)$. Therefore, there are $q$ combinations possible, and $q-2$ lead to successful decoding at the destinations (if $b \notin \{0, 1\}$) and the success probability is $\frac{q-2}{q}$. This probability is zero for the binary field, since in the very particular case of two points in a binary affine geometry, RANC actually reduces to SAF. However, for large $q$, it tends to 1 nearly as fast as its counterpart for RLNC. Finally, note that the decoding of the messages at destination $D_1$, which receives the points $(0|\mathbf{m})$ and $(1-b|b\mathbf{m} + (1-b)\mathbf{n})$, is straightforward (similarly for $D_2$): construct the matrix $\left( \begin{array}{cc|c} 1 & 0 & \mathbf{m} \\ 1 & 1-b & b\mathbf{m} + (1-b)\mathbf{n} \end{array} \right)$, which after Gaussian elimination yields $\left( \begin{array}{cc|c} 1 & 0 & \mathbf{m} \\ 0 & 1 & -\mathbf{m} + \mathbf{n} \end{array} \right)$, and obtain $\mathbf{m}$ and $\mathbf{n}$.

## V. Parameters for error-free networks

### A. General assumptions

In this section, we define, determine, and compare some performance parameters of different matroids, hence leading to a performance comparison of different network protocols. In order to carry out this study, we need to make the following assumption to the model described in Section III. Since our model is noncoherent, it makes the network topology and the statistical dependency amongst packets due to the order of combinations transparent at the message level. Accordingly, we suppose that each destination receives elements chosen independently and uniformly amongst all elements of the flat. This assumption can be viewed as a generalization of the multiplicative matrix channel proposed for RLNC in [18]. Moreover, it is motivated by file dissemination [19] and is similar to the setting in [20]. Note, however, that the study in [20], [21] is based on simple independence assumptions for RLNC and considers delays for the whole set of destinations, while we shall derive fine results for any matroid by viewing each destination separately. Furthermore, we believe this assumption provides a good intuition on how the protocols behave and it allows for the thorough performance study below. The parameters we introduce illustrate the impact of the number of possible combinations offered by different protocols in terms of data rate, delay, and partial decoding.

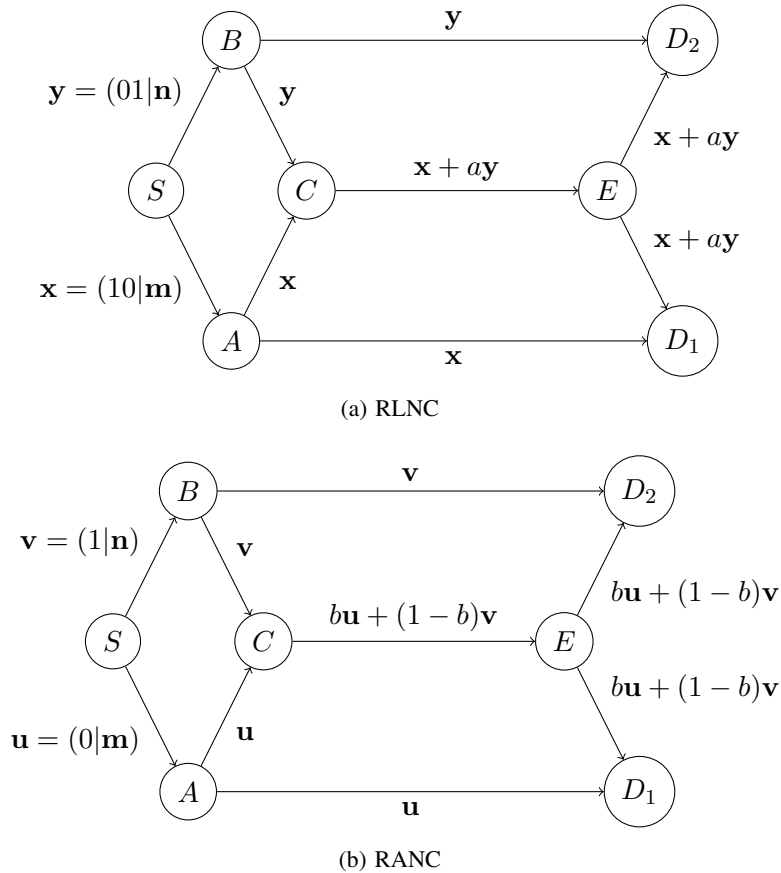

(a) RLNC

(b) RANC

Fig. 1.  Transmitting data on the butterfly network using RLNC or RANC

We comment on the term "error-free" used in the title of this section. Although the destination does not necessarily recover the whole transmitted flat immediately, it keeps receiving elements of the flat and hence it will almost surely be able to reconstruct the whole flat sent by the source. The term error-free indicates that no other flat of the same rank can be reconstructed by the destination, and hence only the message sent by the source can be decoded.

### B. Matroid rate, average delay, and throughput

The data rate of the communication is given by the ratio between the amount of information decoded and the amount of data needed to transmit a flat: $\frac{\log_q A}{nk} = R_{\text{code}}(A, \mathcal{M})R(\mathcal{M}, k)$, where $R_{\text{code}}(A, \mathcal{M}) = \frac{\log_q A}{\log_q N_k}$ can be viewed as the rate of the code formed by all the possible transmitted flats and the *matroid rate* is defined as

$$R(\mathcal{M}, k) = \frac{\log_q N_k}{nk}. \tag{1}$$

We remark that $R_{\text{code}}(A, \mathcal{M})$ only depends on the encoding of the message into a flat, and does not depend on the actual matroid (we only require $N_k \geq A$). Therefore, we only focus on the matroid rate henceforth, which indicates how efficiently a flat of rank $k$ is encoded into a message of $k$ packets. We can further decompose the matroid rate into $R(\mathcal{M}, k) = \frac{\log_q N_k}{k \log_q |E|} \cdot \frac{\log_q |E|}{n}$, where the first ratio is an intrinsic property of the matroid, while the second ratio indicates how efficiently a matroid element is encoded into a packet. Note that the rate is entirely determined by the lattice of flats of $\mathcal{M}$, and does not depend on the cardinalities of flats. Proposition 1 below determines the matroid rates of SAF, RLNC, and RANC.

*Proposition 1 (Matroid rate of SAF, RLNC, and RANC):* The matroid rates of SAF and RLNC are respectively given by

$$1 - \frac{\log_q k}{n} \leq R(\mathcal{S}, k) = \frac{\log_q \binom{q^n}{k}}{nk} \leq 1 - \frac{\log_q k - \log_q e}{n},$$

$$1 - \frac{k}{n} \leq R(\mathcal{L}, k) = \frac{\log_q \begin{bmatrix} n \\ k \end{bmatrix}}{nk} < 1 - \frac{k + \log_{q^k} K_q}{n},$$

$$1 - \frac{k-1}{n} \leq R(\mathcal{A}, k) = \frac{n - k + 1 + \log_q \begin{bmatrix} n \\ k-1 \end{bmatrix}}{nk} < 1 - \frac{k - 1 + \log_{q^k} K_q}{n}.$$

*Proof:* For SAF, the rate is determined by (1) and $N_k = \binom{q^n}{k}$; since $\left(\frac{q^n}{k}\right)^k \leq \binom{q^n}{k} \leq \left(\frac{q^n e}{k}\right)^k$, we obtain the bounds on $R(\mathcal{S}, k)$. For RLNC, we have $N_k = \begin{bmatrix} n \\ k \end{bmatrix}$ and $q^{k(n-k)} \leq \begin{bmatrix} n \\ k \end{bmatrix} < K_q^{-1} q^{k(n-k)}$, as reviewed in Section II-A. For RANC, we have $N_k = q^{n-k+1} \begin{bmatrix} n \\ k-1 \end{bmatrix}$. ∎

RANC allows a gain in terms of rate over RLNC of about one symbol per packet, due to the increase in the number of flats from around $q^{k(n-k)}$ to around $q^{k(n-k+1)}$. This gain follows the fact that RLNC only considers around $q^{-1}$ of all possible packets of length $n$, while RANC considers all possible packets.

According to the assumptions made in Section V-A, the packets arrive at the destination at random. Therefore, the number of packets to be received in order to obtain $k$ independent packets, referred to as the delay of a transmission, is a random variable. Clearly, the minimum delay is exactly $k$, while the maximum delay is unbounded. We hence define the *average delay* of a transmission as the expected number of packets received in order to obtain $k$ independent packets. Clearly, $D(\mathcal{M}, k) \geq k$ for any matroid $\mathcal{M}$. By generalizing the approach typically used to solve the coupon collector problem [22], we obtain for a perfect matroid design where all flats of rank $k$ have cardinality $C_k$ for all $k$,

$$D(\mathcal{M}, k) = \sum_{i=0}^{k-1} \frac{1}{1 - \frac{C_i}{C_k}}. \tag{2}$$

We now determine the value of the average delay for SAF, RLNC, and RANC.

*Proposition 2 (Average delay of SAF, RLNC, and RANC):* The average delays of SAF, RLNC, and RANC are respectively given by

$$k(\log k + \gamma) < D(\mathcal{S}, k) = k \sum_{i=1}^{k} i^{-1} < k(\log k + \gamma) + \frac{1}{2},$$

$$k \leq D(\mathcal{L}, k) = k + \sum_{j=1}^{k-1} \frac{1 - q^{j-k}}{q^j - 1} < k + \frac{q}{(q-1)^2},$$

$$k \leq D(\mathcal{A}, k) = k + \sum_{j=1}^{k-1} \frac{1}{q^j - 1} < k + \frac{q}{(q-1)^2},$$

where $\gamma \approx 0.577$ is Euler's constant.

Proposition 2 indicates that in RLNC and RANC, the expected number of packets needed to decode the subspace completely tends to the rank of the subspace as $q$ tends to infinity. The delay of RANC is very close to that of RLNC since, by (2), the average delay is determined by the cardinality of flats, which only changes from $\frac{q^k-1}{q-1}$ for RLNC to $q^{k-1}$ for RANC.

We now define the *throughput* of a matroid as the ratio between the amount of transmitted information over the amount of data received on average by the destination. In other words, it measures the proportion of useful information in each packet received by the destination. By definition, the throughput is given by

$$T(\mathcal{M}, k) = \frac{\log_q N_k}{n D(\mathcal{M}, k)} = k \frac{R(\mathcal{M}, k)}{D(\mathcal{M}, k)}. \tag{3}$$

This provides an indication on the desirable properties of a matroid for network communications. By (3), a matroid should maintain a low average delay, while trying to maximize its data rate. By (2), minimizing the average delay is equivalent to minimizing the ratio $\frac{C_i}{C_k}$; also, by (1) the matroid rate increases with the number of flats $N_k$. A matroid should hence have a large number of flats, whose cardinalities increase rapidly with their ranks.

Combining the results in Propositions 1 and 2, the throughputs of SAF, RLNC, and RANC are respectively around

$$T(\mathcal{S}, k) \sim \frac{1}{\log k} - \frac{1}{n \log q}, \quad T(\mathcal{L}, k) \sim 1 - \frac{k}{n}, \quad T(\mathcal{A}, k) \sim 1 - \frac{k-1}{n}. \tag{4}$$

By (4), the throughputs of RLNC and RANC are higher for small values of $k$, but decrease linearly with $k$. On the other hand, the throughout of SAF only decreases with the logarithm of $k$, hence this protocol is more appropriate for messages with a large number of packets, which confirms the assumptions made in [2]. The increase in rate and the constant delay between RANC and RLNC lead to a gain in throughput of one symbol per packet in (4).

### C. Number of received independent packets

We now investigate the delay in higher detail by considering the random variable $X_I(\mathcal{M}, k; r)$ given by the number of informative packets received by the destination once $r$ packets have been received. The variable $r - X_I(\mathcal{M}, k; r)$ measures the random redundancy inherent to the noncoherent transmission. Therefore, $\Pr\{X_I(\mathcal{M}, k; r) = l\}$ is equal to the probability $P_I(\mathcal{M}, k; r, l)$ to obtain $l$ independent packets once $r$ packets have been received. An important special case is given by the probability $P_I(\mathcal{M}, k; k, k)$ to receive all the necessary independent packets to reconstruct the flat with minimum delay. Proposition 3 below determines a recursive way of computing $P_I(\mathcal{M}, k; r, l)$.

*Proposition 3 (Probability of independence):* We have $P_I(\mathcal{M}, k; r, 0) = 0$ and for $r \geq 0$,

$$P_I(\mathcal{M}, k; r+1, l+1) = \left(1 - \frac{C_l}{C_k}\right) P_I(\mathcal{M}, k; r, l) + \frac{C_{l+1}}{C_k} P_I(\mathcal{M}, k; r, l+1).$$

In particular, $P_I(\mathcal{M}, k; k, k) = \prod_{i=1}^{k-1} \left(1 - \frac{C_i}{C_k}\right)$.

*Proof:* In order to obtain $l + 1$ independent packets after receiving $r + 1$ packets, one must have received either $l$ or $l+1$ independent packets in the first $r$ received packets. Hence $P_I(\mathcal{M}, k; r+1, l+1) = p_0 P_I(\mathcal{M}, k; r, l) + p_1 P_I(\mathcal{M}, k; r, l+1)$, where $p_0 = \frac{C_k - C_l}{C_k}$ is the probability to receive a packet outside of a flat of rank $l$ and $p_1 = \frac{C_{l+1}}{C_k}$ is the probability to receive a packet inside of a flat of rank $l + 1$. Applying this recursion successively for $l = r$ yields $P_I(\mathcal{M}, k; k, k)$. ∎

We derive closed-form formulas of the probability of independence for SAF, RLNC, and RANC in Proposition 4 below.

*Proposition 4 (Probability of independence for SAF, RLNC, and RANC):* We have for all $l \geq 1$

$$P_I(\mathcal{S}, k; r, l) = \frac{k!}{k^r(k-l)!} \left\{ \begin{matrix} r \\ l \end{matrix} \right\} = \frac{\binom{k}{l}}{k^r} \sum_{j=0}^{l} (-1)^{l-j} \binom{l}{j} j^r,$$

$$P_I(\mathcal{L}, k; r, l) = \frac{\left[ \begin{matrix} k \\ l \end{matrix} \right]}{(q^k-1)^r} \sum_{s=0}^{r-l} (-1)^s \binom{r}{s} \prod_{i=0}^{l-1} (q^{r-s} - q^i),$$

$$P_I(\mathcal{A}, k; r, l) = q^{-(k-1)(r-1)} \left[ \begin{matrix} k-1 \\ l-1 \end{matrix} \right] \prod_{i=0}^{l-2} (q^{r-1} - q^i),$$

where $\left\{ \begin{smallmatrix} r \\ l \end{smallmatrix} \right\}$ is a Stirling number of the second kind [23]. In particular,

$$\sqrt{2\pi k}\, e^{-k+\frac{1}{12k+1}} < P_I(\mathcal{S}, k; k, k) = \frac{k!}{k^k} < \sqrt{2\pi k}\, e^{-k+\frac{1}{12k}},$$

$$K_q < P_I(\mathcal{L}, k; k, k) = \prod_{i=1}^{k-1} \left( 1 - \frac{q^i - 1}{q^k - 1} \right) \leq 1,$$

$$K_q < P_I(\mathcal{A}, k; r, l) = q^{-(k-1)(r-1)} \left[ \begin{matrix} k-1 \\ l-1 \end{matrix} \right] \prod_{i=0}^{l-2} (q^{r-1} - q^i) \leq 1.$$

is the probability that all $k$ independent packets are received with minimum delay.

*Proof:* For SAF, $P_I(\mathcal{S}, k; r, l) = \frac{S_I(r,l)}{k^r}$, where $S_I(r, l)$ is the number of words of length $r$ with $l$ distinct symbols from an alphabet of size $k$. Any word with $l$ distinct symbols can be put in correspondence with the partition of $[r]$ into $l$ cells, where each cell contains the positions of a given symbol in the word. By definition of the Stirling numbers, there are $\left\{ \begin{smallmatrix} r \\ l \end{smallmatrix} \right\}$ such partitions. Also, once the partition is fixed, there are $k(k-1)\cdots(k-l+1)$ choices for the symbols. Combining, we obtain the formula for $P_I(\mathcal{S}, k; r, l)$. For $r = l = k$, we obtain $P_I(\mathcal{S}, k; k, k) = \frac{k!}{k^k}$, which combined with the refinement of Stirling's formula in [24] yields the upper bound.

For RLNC, we have $P_I(\mathcal{L}, k; r, l) = \frac{R_I(r,l)}{\left[ \begin{smallmatrix} k \\ 1 \end{smallmatrix} \right]^r}$, where $R_I(r, l)$ is the number of matrices in $\mathrm{GF}(q)^{k \times r}$ with rank $l$ such that all the columns are nonzero and the leading nonzero coefficient is equal to 1. The number of matrices with rank $l$ and $s$ nonzero columns is hence given by $\binom{r}{s}(q-1)^s R_I(s, l)$. Also, the number of matrices in $\mathrm{GF}(q)^{k \times r}$ with rank $l$ is given by $\left[ \begin{smallmatrix} k \\ l \end{smallmatrix} \right] \prod_{i=0}^{l-1}(q^r - q^i)$ [5], [7]. Summing all matrices with rank $l$ and $s$ nonzero columns for $l \leq s \leq r$, we obtain

$$\sum_{s=l}^{r} \binom{r}{s}(q-1)^s R_I(s, l) = \left[ \begin{matrix} k \\ l \end{matrix} \right] \prod_{i=0}^{l-1}(q^r - q^i).$$

By applying the reverse binomial transform [25], we obtain the formula for $P_I(\mathcal{L}, k; r, l)$.

For RANC, we have $P_I(\mathcal{A}, k; r, l) = \frac{A_I(r,l)}{q^{r(k-1)}}$, where $A_I(r, l)$ is the number of collections of points $\mathbf{v}_0, \mathbf{v}_1, \ldots, \mathbf{v}_{r-1} \in \mathrm{GF}(q)^n$ in a flat of rank $k$ with exactly $l$ affinely independent points. We have $\text{rk}(\mathbf{v}_0, \mathbf{v}_1, \ldots, \mathbf{v}_{r-1}) = \text{rank}(\mathbf{1}|\mathbf{V}) = 1 + \text{rank}(\mathbf{W})$, where $\mathbf{W} = ((\mathbf{v}_1 - \mathbf{v}_0)^T, (\mathbf{v}_2 - \mathbf{v}_0)^T, \ldots, (\mathbf{v}_{r-1} - \mathbf{v}_0)^T)^T \in \text{GF}(q)^{(r-1)\times n}$ is a matrix with rank $l-1$ whose rows belong to a linear subspace of dimension $k-1$. There are $q^{k-1}$ choices for $\mathbf{v}_0$ and $\begin{bmatrix} k-1 \\ l-1 \end{bmatrix} \prod_{i=0}^{l-2}(q^{r-1} - q^i)$ choices for $\mathbf{W}$, and hence $A_I(r,l) = q^{k-1}\begin{bmatrix} k-1 \\ l-1 \end{bmatrix}\prod_{i=0}^{l-2}(q^{r-1} - q^i)$ which leads to the result for $P_I(\mathcal{A}, k; r, l)$. ∎

We now investigate the moments of the probability distribution $P_I(\mathcal{M}, k; r, l)$, in particular the expectation $E_I(\mathcal{M}, k; r)$ and the variance $V_I(\mathcal{M}, k; r)$. We clearly have $E_I(\mathcal{M}, k; r) \leq \min\{k, r\}$, $E_I(\mathcal{M}, k; r) \leq E_I(\mathcal{M}, k; r+1) \leq E_I(\mathcal{M}, k; r) + 1$, and $\lim_{r\to\infty} E_I(\mathcal{M}, k; r) = k$ and $\lim_{r\to\infty} V_I(\mathcal{M}, k; r) = 0$. Proposition 5 below determines or bounds the expectation and the variance for SAF, RLNC, and RANC.

*Proposition 5 (Average number of independent elements in SAF, RLNC, and RANC):* For all $k$ and $r$,

$$E_I(\mathcal{S}, k; r) = k\left[1 - \left(1 - \frac{1}{k}\right)^r\right] \sim k(1 - e^{-\frac{r}{k}}). \tag{5}$$

Also, the variance is given by

$$V_I(\mathcal{S}, k; r) = k\left[\left(1 - \frac{1}{k}\right)^r - \left(1 - \frac{2}{k}\right)^r\right] + k^2\left[\left(1 - \frac{2}{k}\right)^r - \left(1 - \frac{1}{k}\right)^{2r}\right] \sim ke^{-\frac{r}{k}}(1 - e^{-\frac{r}{k}}).$$

For RLNC and RANC, we have $P_I(\mathcal{L}, k; r, r) > K_q$ and $P_I(\mathcal{A}, k; r, r) > K_q$ for all $r \leq k$ by Proposition 4. Therefore, $E_I(\mathcal{L}, k; r) > K_q\min\{r, k\}$ and $E_I(\mathcal{A}, k; r) > K_q\min\{r, k\}$ for all $r$, and accordingly, the variance tends to 0 with the field size.

*Proof:* Let $S_I(r, l)$ denote the number of words of length $r$ with $l$ distinct symbols, then $P_I(\mathcal{S}, k; r, l) = k^{-r}S_I(r, l)$. Consider the bipartite graph on $[k]^r$ and $\mathcal{P}([k], a)$ $(1 \leq a \leq k)$, where two vertices are adjacent if and only if there exist $a$ symbols of the word in $[k]^r$ equal to the $a$ elements in $[k]$. Let us count the number of edges in this graph in two different ways. First, there are $\binom{l}{a}$ edges adjacent to a word in $[k]^r$ with $l$ different coefficients, hence there are $\sum_{l=a}^{k}\binom{l}{a}S_I(r, l)$ edges in the graph. Second, by the inclusion-exclusion principle, we obtain that each subset of $[k]$ with cardinality $a$ appears in exactly $\sum_{i=0}^{a}(-1)^i\binom{a}{i}(k-i)^r$ words in $[k]^r$. Therefore,

$$\sum_{l=a}^{k}\binom{l}{a}S_I(r, l) = \binom{k}{a}\sum_{i=0}^{a}(-1)^i\binom{a}{i}(k-i)^r,$$

which yields (5) for $a = 1$. Using this identity for $a = 2$ and combining, we also obtain the variance. ∎

In particular, for $r = k$, (5) indicates that only around $1 - e^{-1} \approx 0.632$ of the first $k$ received packets are independent on average. On the other hand, for RLNC and RANC, the average number of independent packets tends to the optimal with the field size.

The expected number of independent elements for RANC and SAF determined or bounded above is illustrated in Figure 2 for $q = 2^8$, $n = 20$, $k = 10$, and $1 \leq r \leq 30$. For SAF, the exponential pattern determined in Proposition 5 is clearly displayed, while RANC is close to optimal for practical values of $q$. For $k = 10$, Proposition 2 indicates that the average delay is given by $D(\mathcal{S}(2^8, 20), 10) \approx 29.3$ for SAF while $D(\mathcal{A}(2^8, 20), 10) \approx 10.004$ for RANC.

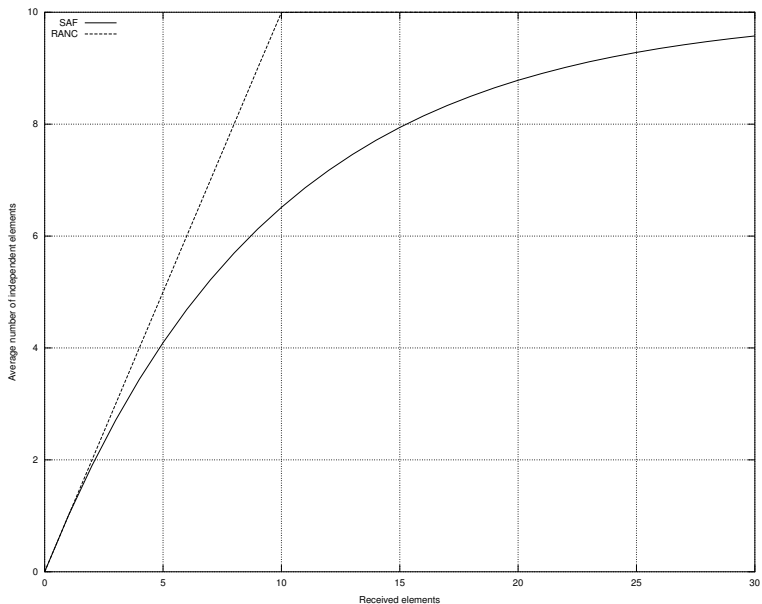

Fig. 2. Expected number of independent elements as a function of the number of received elements for $k = 10$ transmitted elements

### D. Partial decoding

The model introduced in Section III and the parameters determined so far assume the destination waits to receive $k$ independent packets in order to begin the decoding procedure. However, the destination may choose to operate partial decoding on a fraction of $k$ independent packets. The problem of partial decoding is hence as follows: Suppose that less than $k$ independent packets have been received, how many packets do we expect to decode? We remark that this problem is intrinsic to the matroid, and does not depend on the assumptions on the model made in the introduction of Section V.

The destination can perform partial decoding if it knows a way of recovering the messages in all the elements contained in the flat it has received which were originally transmitted by the source. This is equivalent to transmitting elements of a *canonical basis*, defined as follows. A basis $B(f)$ of a flat $f$ is a canonical basis if the elements of $B(f)$ can be discriminated from the other elements of $f$ and if the information carried by any element of $B(f)$ can be decoded using this element only. The first property guarantees that the destination receiving a subflat $g$ of the transmitted flat $f$ will be able to determine all the elements transmitted by the source which lie in $g$, while the second property guarantees that the destination will decode the information carried by these elements. In other words, a canonical basis is a systematic encoding of the message, and the destination wishes to retrieve the systematic part of the elements of the canonical basis. It is still unknown which flats of a matroid have a canonical basis; however, it can be easily shown that the liftings associated to SAF, RLNC, and RANC have a canonical basis given by the set of rows of the matrix.

We illustrate partial decoding using the linear lifting for RLNC. Suppose three messages $\mathbf{m}$, $\mathbf{n}$, and $\mathbf{p}$ over $\mathrm{GF}(q)$ are to be sent using RLNC. After linear lifting, the source transmits the following three vectors: $(100|\mathbf{m})$, $(010|\mathbf{n})$, and $(001|\mathbf{p})$, which form the canonical basis of the transmitted flat. Suppose the destination first receives the vector $(1a0|\mathbf{m} + a\mathbf{n})$, which does not belong to the canonical basis and hence cannot be decoded. Following, suppose the destination then receives $(1b0|\mathbf{m} + b\mathbf{n})$. Two members of the canonical basis clearly belong to the closure of the received vectors, therefore the destination can decode the two messages $\mathbf{m}$ and $\mathbf{n}$ before receiving the entire flat.

Let $P_D(\mathcal{M}, k; l, d)$ be the probability to decode $d$ elements after receiving $l$ independent elements. Provided a canonical basis exists, decoding $d$ elements is equivalent to receiving a flat containing $d$ members of the canonical basis of the transmitted flat. We determine $P_D(\mathcal{M}, k; l, d)$ under certain assumptions on the matroid, which are satisfied by $\mathcal{L}$, $\mathcal{S}$, and $\mathcal{A}$.

*Proposition 6 (Probability of partial decoding):* Suppose the transmitted flat of rank $k$ contains $G(l, k)$ flats of rank $l$ for all $0 \leq l \leq k$. Furthermore, assume that for all $a \leq l \leq k$, any flat with rank $a$ is contained in $F(a, l)$ flats of rank $l$, all of them being contained in the transmitted flat. We then have

$$P_D(\mathcal{M}, k; l, d) = \frac{\binom{k}{d}}{G(l, k)} \sum_{a=d}^{l} (-1)^{a+d} \binom{k-d}{a-d} F(a, l).$$

*Proof:* The probability is given by $P_D(\mathcal{M}, k; l, d) = \frac{N_D(l, d)}{G(l, k)}$, where $N_D(l, d)$ is the number of flats of rank $l$ that are contained within the transmitted flat $f$ and which have $d$ decodable elements. We now determine the value of $N_D(l, d)$.

First, the set of flats of rank $l$ with $l$ decodable elements is given by $\{\mathrm{cl}(X) : |X| = l, X \subseteq B(f)\}$, and hence $N_D(l, l) = \binom{k}{l}$. Now consider the bipartite graph on the set of subflats of $f$ of rank $l$ and the set of flats of rank $a$ with $a$ decodable elements ($a \leq d \leq l$), where two vertices $f_l \in \mathcal{F}_l, f_a \in \mathcal{F}_a$ are adjacent if and only if $f_a \subseteq f_l$. We now count the edges in this graph in two ways. Since there are

$F(a, l)$ edges adjacent to any flat of rank $a$, the number of edges is given by $\binom{k}{a}F(a, l)$. Also, there are $\binom{d}{a}$ edges adjacent to any flat of rank $l$ with $d$ decodable elements, hence there are $\sum_{d=a}^{l}\binom{d}{a}N_D(l, d)$ edges. Combining, we obtain

$$\sum_{d=a}^{l}\binom{d}{a}N_D(l, d) = \binom{k}{a}F(a, l). \tag{6}$$

Denoting $\mathbf{n} = (N_D(l, a), N_D(l, a+1), \ldots, N_D(l, l))$, $\mathbf{v} = (\binom{k}{a}F(a, l), \binom{k}{a+1}F(a+1, l), \ldots, \binom{k}{l}F(l, l))$, (6) becomes $\mathbf{nL} = \mathbf{v}$, where $\mathbf{L} = (l_{d,a})$ is a Pascal matrix: $l_{d,a} = \binom{d}{a}$ [26]. Since $\mathbf{L}^{-1} = (m_{a,d})$ has $m_{a,d} = (-1)^{d+a}\binom{a}{d}$, we obtain the formula for $P_D(\mathcal{M}, k; l, d)$. ∎

We remark that (6) provides the binomial moments of the $P_D(\mathcal{M}, k; l, d)$ distribution. In particular, the expectation $E_D(\mathcal{M}, k; l)$ and the variance $V_D(\mathcal{M}, k; l)$ are respectively given by

$$E_D(\mathcal{M}, k; l) = k\frac{F(1, l)}{G(l, k)},$$
$$V_D(\mathcal{M}, k; l) = k\frac{(k-1)F(2, l) + F(1, l)}{G(l, k)} - k^2\frac{F(1, l)^2}{G(l, k)^2}.$$

*Corollary 1 (Probability of partial decoding for SAF, RLNC, and RANC):* For SAF, $P_D(\mathcal{S}, k; l, d) = \delta_{l-d}$ for all $l, d$, and hence $E_D(\mathcal{S}, k; l) = l$ and $V_D(\mathcal{S}, k; l) = 0$ for all $l$. For RLNC, we have $E_D(\mathcal{L}, k; l) = k\frac{q^l - 1}{q^k - 1} < kq^{l-k}$ for all $l$. For RANC, we have $E_D(\mathcal{A}, k; l) = kq^{l-k}$ for all $l$.

*Proof:* For SAF, all elements are decodable and $P_D(\mathcal{S}, k; l, d) = \delta_{l-d}$ for all $l, d$. This can also be demonstrated via Proposition 6, where $F(a, l) = \binom{k-a}{l-a}$ and $G(l, k) = \binom{k}{l}$. For RLNC, we have $F(a, l) = \begin{bmatrix} k-a \\ l-a \end{bmatrix}$ and $G(l, k) = \begin{bmatrix} k \\ l \end{bmatrix}$ by [27, Lemma 2], which yields $E_D(\mathcal{L}, k; l) = k\frac{\begin{bmatrix} k-1 \\ l-1 \end{bmatrix}}{\begin{bmatrix} k \\ l \end{bmatrix}}$. For RANC, we have $F(0, l) = q^{k-l}\begin{bmatrix} k-1 \\ l-1 \end{bmatrix}$, $F(a, l) = \begin{bmatrix} k-a \\ l-a \end{bmatrix}$ for $a > 0$, and $G(l, k) = q^{k-l}\begin{bmatrix} k-1 \\ l-1 \end{bmatrix}$. ∎

We remark that $E_D(\mathcal{A}, k; k-1) = kq^{-1}$ by Corollary 1; hence for $q = 2^8$ only 0.39% of the packets can be decoded before receiving all the packets. Therefore, for practical values of the field size, RANC (and also RLNC) hardly offers any opportunity of partial decoding.

Finally, let $P_T(\mathcal{M}, k; r, d)$ be the probability to decode $d$ packets given that $r$ packets (not necessarily independent) have been received. Clearly, $P_T(\mathcal{M}, k; r, d) = \sum_{l=0}^{r} P_I(\mathcal{M}, k; r, l)P_D(\mathcal{M}, k; l, d)$, and hence we can regroup the results above to determine the probability $P_T(\mathcal{M}, k; r, d)$. For SAF, by Corollary 1, the expected number of decodable packets is given by $E_T(\mathcal{S}, k; r) = E_I(\mathcal{S}, k; r) \sim k(1 - e^{-\frac{r}{k}})$. For RLNC and RANC, we respectively have $E_T(\mathcal{L}, k; r) \leq E_D(\mathcal{L}, k; r) < kq^{r-k}$ and $E_T(\mathcal{A}, k; r) \leq kq^{r-k}$. In particular, $E_T(\mathcal{A}, k; k-1) \leq kq^{-1}$, hence only $q^{-1}$ of the packets can be partially decoded before receiving $k$ packets. RANC then follows a zero-one behavior: before receiving $k$ packets, no decoding is possible; once $k$ packets are received, they are independent with high probability and the whole message can be decoded. This behavior is illustrated in Figure 3, where the values of $E_T(\mathcal{S}, k; l)$ and $E_T(\mathcal{A}, k; l)$ for $q = 2^8$ are displayed for $k = 10$.

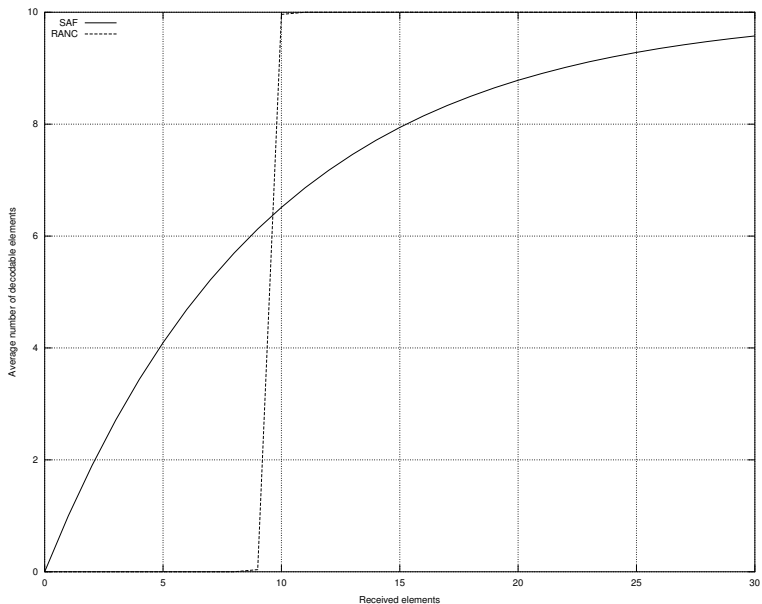

Fig. 3. Expected number of decodable elements as a function of the number of received elements for $k = 10$ transmitted elements

## VI. MATROID ERROR-CORRECTING CODES

### A. Operator channel and metrics for error correction

In Section III, we modeled data communication through a network as the transmission of a flat of a matroid. However, the model in Section III did not take into account the possible alterations undergone by the message during its transmission through the network. These alterations due to the network—packet losses, injections, errors, etc.—modify not only the packets but also the flat being transmitted. A flat $f \in \mathcal{F}_k$ can be modified in two ways: a *deletion* turns $f$ into a proper subflat of rank $k - 1$, while an *insertion* turns $f$ into a proper superflat of rank $k + 1$. A deletion (an insertion, respectively) is hence equivalent to moving one step down (up, respectively) the lattice of flats. Any flat $f$ can be turned into any other flat $g$ via a sequence of insertions and deletions. The terms "insertion" and "deletion" were first introduced in [4] for RLNC.

Proposition 7 below proves that the shortest way to modify one flat into another is to perform all the insertions first, and then all the deletions. This can be intuitively explained as follows. By performing the insertions first, the network works in larger flats with a much higher cardinality. This large number of combinations allows to produce bases with 'distant' elements to the original flat, and hence drift away from the original flat without taking steps on the lattice. It then suffices to go back down by deleting some elements. On the other hand, performing the deletions first implies to work in smaller flats, which hinders the message from drifting away from the original flat.

*Proposition 7:* For any pair of flats $f, g$ of a matroid, the union-path $U(f, g)$, defined as starting from $f$, going up the lattice of flats to $\mathrm{cl}(f \cup g)$, and then going back down to $g$, is a shortest path between $f$ and $g$. Therefore, the shortest path distance $s(f, g)$ between $f$ and $g$ is given by $s(f, g) = 2\mathrm{rk}(f \cup g) - \mathrm{rk}(f) - \mathrm{rk}(g)$.

*Proof:* Without loss of generality, suppose $\mathrm{rk}(f) \geq \mathrm{rk}(g)$. We shall prove the claim by induction on $s(f, g)$. First, the cases $s(f, g) = 0$ and $s(f, g) = 1$ are trivial. Also, if $s(f, g) = 2$, then either $g \subset f$ and the union-path is the only path of length 2, or $g \nsubseteq f$ and the only path of length 2 distinct from the union-path is the intersection-path $\{f, f \cap g, g\}$. The intersection-path has length $\mathrm{rk}(f) + \mathrm{rk}(g) - 2\mathrm{rk}(f \cap g)$; however, $2\mathrm{rk}(f \cup g) - \mathrm{rk}(f) - \mathrm{rk}(g) \leq \mathrm{rk}(f) + \mathrm{rk}(g) - 2\mathrm{rk}(f \cap g)$ by the submodular inequality, and hence the union-path is no longer than the intersection-path. Therefore, the union-path is among the shortest paths for $s(f, g) = 2$.

Now suppose the claim is true for all pairs of flats with a shortest path of length no more than $d$, and consider $f$ and $g$ such that $s(f, g) = d + 1$. Let $\{f, p_1, \ldots, p_d, g\}$ be a shortest path between $f$ and $g$. Since $s(p_{d-1}, g) = 2$, we assume $g \subset p_d$, without loss by the discussion above. However, $s(f, p_d) = d$ and hence $U(f, p_d)$ is a shortest path between $f$ and $p_d$. Therefore, $\{U(f, p_d)\} \cup \{g\}$ is a shortest path between $f$ and $g$ which first goes up the lattice and then down, and hence is equal to $U(f, g)$. ∎

We model data transmission through a faulty network as an operator channel, where the source transmits a flat $f \in \mathcal{F}$ and the destination obtains another flat $g \in \mathcal{F}$, which is obtained after $\delta$ insertions and $\epsilon$ deletions, where $\delta = \mathrm{rk}(f \cup g) - \mathrm{rk}(f)$ and $\epsilon = \mathrm{rk}(f \cup g) - \mathrm{rk}(g)$, respectively. Accordingly, we define the lattice distance between $f$ and $g$ as

$$
\begin{aligned}
d_{\mathrm{L}}(f, g) &= \delta + \epsilon \\
&= 2\mathrm{rk}(f \cup g) - \mathrm{rk}(f) - \mathrm{rk}(g) & (7) \\
&\leq \mathrm{rk}(f \cup g) - \mathrm{rk}(f \cap g) & (8) \\
&\leq \mathrm{rk}(f) + \mathrm{rk}(g) - 2\mathrm{rk}(f \cap g), & (9)
\end{aligned}
$$

where (8) and (9) follow the submodular inequality.

For SAF, the lattice distance between two subsets is their Hamming distance; for RLNC, the lattice distance between two linear subspaces is their subspace distance. We remark that for both RLNC and SAF, we have equality in (8) and (9) for all flats. This however does not hold for all matroids; those which satisfy this property can all be expressed as direct sums of free matroids and projective geometries [3, Proposition 6.9.1].

Let us illustrate these remarks with the affine geometry. Two parallel hyperplanes $f$ and $g$ in $\mathcal{A}(q, n)$ are at lattice distance 2, while (8) and (9) yield $n + 1$ and $2n$, respectively. Furthermore, by considering $f$, $g$, and the whole space, it can be easily shown that the right hand sides of (8) and (9) violate the triangular inequality. This example also illustrates how the lattice distance expresses the minimum number of operations required to change one flat into the other. In our example, changing $f$ into $g$ takes only two operations: first insert an element not belonging to $f$ to obtain the whole space, which has a basis given by $r$ elements of $g$ and one outside of $g$; then delete the latter element to obtain $g$.

The modified lattice distance

$$
\begin{aligned}
d_{\mathrm{M}}(f, g) &= \max\{\delta, \epsilon\} \\
&= \mathrm{rk}(f \cup g) - \min\{\mathrm{rk}(f), \mathrm{rk}(g)\} \tag{10} \\
&\leq \max\{\mathrm{rk}(f), \mathrm{rk}(g)\} - \mathrm{rk}(f \cap g) \tag{11}
\end{aligned}
$$

coincides with the modified Hamming metric for SAF [2] and with the injection distance for RLNC [4]. Similarly to the lattice distance, equality holds in (11) for SAF and RLNC; however, in the example above inequality is strict and the triangular inequality is violated by the right hand side.

We remark that both distances only depend on the lattice of flats of the matroid. However, for any non-simple matroid, there exists a simple matroid with the same lattice of flats. Therefore, our assumption in Section III of considering simple matroids only does not lead to any loss in generality.

Corollary 2 ensures that the distances defined above are metrics. Therefore, error correction for random network communications can be viewed as a coding theory problem, where the codewords are flats of a matroid associated to the network protocol and the distance between two flats is either the lattice distance or the modified lattice distance.

*Corollary 2:* For any simple matroid with rank $r$, the lattice distance and the modified lattice distance associated to that matroid are metrics which take integer values between 0 and $r$.

*Proof:* The lattice distance is a metric according to Proposition 7. For the modified lattice distance, by (7) and (10) we have $d_{\mathrm{M}}(f, g) = \frac{1}{2}d_{\mathrm{L}}(f, g) + \frac{1}{2}|\mathrm{rk}(f) - \mathrm{rk}(g)|$ for all flats $f, g \in \mathcal{F}$, and hence we

easily obtain that $d_{\mathrm{M}}$ is also a metric. It is clear that these metrics only have integer values between $0$ and $r$. ∎

### B. Matroid codes

For any simple matroid $\mathcal{M}$, we define a *matroid code* as a nonempty set of flats of a matroid with the same rank, or equivalently as a subset of $\mathcal{F}_k$. The minimum lattice (modified lattice, respectively) distance of a matroid code is given by the minimum distance between two pairs of distinct codewords. By (7) and (10), the minimum lattice distance of a matroid code is twice its minimum modified lattice distance. All the other classical parameters of a code, such as the error correction capability, the covering radius, the diameter, etc. can be similarly defined.

We now derive bounds on matroid codes which are natural generalizations of the bounds derived for constant-dimension codes and constant-weight codes. Indeed, the latter classes of codes are so well structured and can be easily bounded because they actually are matroid codes. In other words, by studying matroid codes, we investigate some core properties of these classes of codes. We denote the maximum cardinality of a matroid code on the flats of rank $k$ of $\mathcal{M}$ with *minimum modified lattice distance $d$* (and hence minimum lattice distance $2d$) as $A(\mathcal{M}, k, d)$. Denoting the rank of $\mathcal{M}$ as $r$, we first remark that $A(\mathcal{M}, k, 1) = N_k$ for all $0 \leq k \leq r$. Also, since $d_{\mathrm{M}}(f, g) \leq \min\{k, r - k\}$ for all $f, g \in \mathcal{F}_k$, we shall use the following convention: $A(\mathcal{M}, k, d) = 1$ for all $d > \min\{k, r - k\}$.

Johnson bounds have first been derived for constant-weight codes [28] and have been adapted to constant-dimension codes in [10], [12]. Proposition 8 below generalizes these bounds to the case of matroid codes by restricting to a submatroid of inferior rank in two ways. First, for any $e \in E$, the *contraction* of $e$ from $\mathcal{M}$, denoted as $\mathcal{M}/e$, is the simple matroid with set of flats $\mathcal{F}(\mathcal{M}/e) = \{f \in \mathcal{F} : e \in f\}$ [3, Chapter 3]. Note that the matroid $\mathcal{M}/e$ has rank $r - 1$ and for all $f, g \in \mathcal{F}(\mathcal{M}/e)$, $\mathrm{rk}_{\mathcal{M}/e}(f) = \mathrm{rk}_{\mathcal{M}}(f) - 1$ and hence $d_{\mathrm{M}, \mathcal{M}/e}(f, g) = d_{\mathrm{M}, \mathcal{M}}(f, g)$. Second, for any hyperplane $h \in \mathcal{F}_{r-1}$, the *restriction* of $\mathcal{M}$ to $h$, denoted as $\mathcal{M}|h$, is the simple matroid with set of flats $\{f \in \mathcal{F} : f \subseteq h\}$ [3, Section 1.3]. For any flats $f, g \subseteq h$, $\mathrm{rk}_{\mathcal{M}|h}(f) = \mathrm{rk}_{\mathcal{M}}(f)$ and hence $d_{\mathrm{M}, \mathcal{M}|h}(f, g) = d_{\mathrm{M}, \mathcal{M}}(f, g)$.

*Proposition 8 (Johnson bound):* For all $\mathcal{M}$ and $0 \leq k \leq r$, denote the minimum cardinality of a flat of rank $k$ and the minimum number of hyperplanes containing a given flat of rank $k$ as $c_k$ and $h_k$, respectively. Then there exist $e \in E$ and $h \in \mathcal{F}_{r-1}$ such that

$$A(\mathcal{M}, k, d) \leq \frac{N_1}{c_k} A(\mathcal{M}/e, k - 1, d), \tag{12}$$

$$A(\mathcal{M}, k, d) \leq \frac{N_{r-1}}{h_k} A(\mathcal{M}|h, k, d). \tag{13}$$

*Proof:* We only prove (12), the proof of (13) being similar. For all $f \in \mathcal{F}$ and $e \in E$, let $\chi(e, f) = 1$ if $e \in f$ and $\chi(e, f) = 0$ otherwise. Let $\mathcal{C}$ be a code on the flats of $\mathcal{M}$ with rank $k$, minimum distance $d$, and cardinality $A(\mathcal{M}, k, d)$. Then for all $e \in E$, $\mathcal{C} \cap (\mathcal{F}/e)$ can be viewed a code on the flats of $\mathcal{M}/e$ with rank $k - 1$, minimum distance at least $d$, and cardinality $\sum_{f \in \mathcal{C}} \chi(e, f) \le A(\mathcal{M}/e, k-1, d)$. Conversely, we have $\sum_{e \in E} \chi(e, f) = |f| \ge c_k$ for all $f \in \mathcal{C}$. Combining, we obtain that there exists an element $e' \in E$ for which $|E| A(\mathcal{M}/e', k-1, d) \ge \sum_{e \in E} \sum_{f \in \mathcal{C}} \chi(e, f) \ge c_k A(\mathcal{M}, k, d)$. ∎

Proposition 9 below is a generalization of the Singleton bound for constant-dimension codes derived in [1].

*Proposition 9 (Singleton bound):* For all $\mathcal{M}$, $0 \le k \le r$, and any element $e \in E$, we have $A(\mathcal{M}, k, d) \le A(\mathcal{M}/e, k, d-1)$ and hence $A(\mathcal{M}, k, d) \le \min_{g \in \mathcal{F}_{d-1}} |\{f \in \mathcal{F}_{k+d-1} : g \subseteq f\}|$.

*Proof:* Let $\mathcal{C}$ be a code on $\mathcal{F}_k$ with minimum distance $d$ and cardinality $A(\mathcal{M}, k, d)$. For any $f \in \mathcal{C}$, we define the puncturing $H_e(f)$ as a flat in $\mathcal{F}_k(\mathcal{M}/e)$ containing $f$. Then by the lengths of the shortest paths on the lattice of flats of $\mathcal{M}$, we have $d_{\mathrm{L}}(H_e(f), H_e(g)) \ge d_{\mathrm{L}}(f, g) - 2$ and hence $d_{\mathrm{M}}(H_e(f), H_e(g)) \ge d_{\mathrm{M}}(f, g) - 1$ for all $f, g \in \mathcal{C}$. Therefore, $\{H_e(f) : f \in \mathcal{C}\}$ is a code on the flats of $\mathcal{M}/e$ with rank $k$, minimum distance at least $d - 1$, and cardinality $A(\mathcal{M}, k, d) \le A(\mathcal{M}/e, k, d-1)$. Applying this bound recursively yields the second upper bound. ∎

We finish this section by noting that the concept of lifting, used to construct good matroid codes for RLNC [1], [13] and SAF [2], could be generalized for any matroid. However, as shown in the case of SAF [2], these codes may not be optimal, hence we shall not develop this idea any further.

### C. Matroid codes for the affine geometry

We are now interested in matroid codes on affine subspaces. By definition, we have $A(\mathcal{A}, k, 1) = N_k = q^{n-k+1} {n \brack k-1} \ge q^{k(n-k+1)}$ for all $0 \le k \le n + 1$. Since the affine geometry $\mathcal{A}(q, n)$ is a submatroid of the projective geometry $\mathcal{L}(q, n+1)$, the upper bound on codes on linear subspaces reviewed in Section II-B yields

$$A(\mathcal{A}(q, n), k, d) \le A(\mathcal{L}(q, n+1), k, d) < K_q^{-1} q^{\min\{(n-k+1)(k-d+1), k(n-k-d+2)\}}. \qquad (14)$$

As we shall see later, the upper bound on $A(\mathcal{A}(q, n), k, d)$ in (14) is tight up to a scalar. However, we refine this bound below by applying the Johnson bounds derived in Proposition 8 to codes on affine subspaces.

*Proposition 10 (Bounds on codes on affine subspaces):* For all $2 \le k \le n-1$ and $2 \le d \le \min\{k, n-$

$k + 1\}$, we have

$$
\begin{align}
A(\mathcal{A}(q,n),k,d) &\leq q^{n-k+1} A(\mathcal{L}(q,n),k-1,d), \tag{15}\\
A(\mathcal{A}(q,n),k,d) &\leq \left\lfloor q \frac{q^n - 1}{q^{n-k+1} - 1} A(\mathcal{A}(q,n-1),k,d) \right\rfloor \tag{16}\\
&\leq \left\lfloor q \frac{q^n - 1}{q^{n-k+1} - 1} \left\lfloor q \frac{q^{n-1} - 1}{q^{n-k} - 1} \cdots \left\lfloor q \frac{q^{k+d-1} - 1}{q^d - 1} \right\rfloor \cdots \right\rfloor \right\rfloor.
\end{align}
$$

*Proof:* For any $e \in \mathrm{GF}(q)^n$ and $h \in \mathcal{F}_n(\mathcal{A}(q,n))$, $\mathcal{A}(q,n)/e$ and $\mathcal{A}(q,n)|h$ are isomorphic to $\mathcal{L}(q,n-1)$ and $\mathcal{A}(q,n-1)$, respectively [3, Proposition 6.2.5]. Also, every flat with rank $k$ of $\mathcal{A}(q,n)$ contains $q^{k-1}$ elements and is contained in $\begin{bmatrix} n-k+1 \\ 1 \end{bmatrix}$ hyperplanes. Applying Proposition 9 and (12) and (13) in Proposition 8 to $\mathcal{A}(q,n)$ hence leads to (15) and (16), respectively. Finally, applying (16) recursively yields the last upper bound. ∎

We remark that the first Johnson bound in (15) is good for $2k \leq n+1$, while the second Johnson bound in (16) is good for $2k \geq n+1$. Also, the bounds obtained by applying the Singleton bound in Proposition 9 are looser than (15), and are hence omitted.

Recall that for any $\mathbf{M} \in \mathrm{GF}(q)^{k \times (n-k+1)}$, the affine lifting of $\mathbf{M}$, hereby denoted as $I_{\mathcal{A}}(\mathbf{M}) \in \mathcal{F}_k$, is the closure of the rows of $(\mathbf{I}'_k | \mathbf{M})$, where $\mathbf{I}'_k = (\mathbf{0}|\mathbf{I}_{k-1})^T \in \mathrm{GF}(q)^{k \times (k-1)}$. Remark that $\mathrm{rk}(I_{\mathcal{A}}(\mathbf{M})) = \mathrm{rank}(\mathbf{1}|\mathbf{I}'_k|\mathbf{M}) = k$ for all $\mathbf{M} \in \mathrm{GF}(q)^{k \times (n-k+1)}$, hence $I_{\mathcal{A}}$ indeed maps $\mathrm{GF}(q)^{k \times (n-k+1)}$ to $\mathcal{F}_k$. Proposition 11 below shows that the affine lifting preserves the distance.

*Proposition 11 (Affine lifting):* For any $\mathbf{M}, \mathbf{N} \in \mathrm{GF}(q)^{k \times (n-k+1)}$, we have $d_{\mathrm{M}}(I_{\mathcal{A}}(\mathbf{M}), I_{\mathcal{A}}(\mathbf{N})) = d_{\mathrm{R}}(\mathbf{M}, \mathbf{N})$.

*Proof:* We have

$$
\mathrm{rk}(I_{\mathcal{A}}(\mathbf{M}) \cup I_{\mathcal{A}}(\mathbf{N})) = \mathrm{rank}\left( \begin{array}{c|c|c} \mathbf{1} & \mathbf{I}'_k & \mathbf{M} \\ \hline \mathbf{1} & \mathbf{I}'_k & \mathbf{N} \end{array} \right) = \mathrm{rank}\left( \begin{array}{c|c|c} \mathbf{1} & \mathbf{I}'_k & \mathbf{M} \\ \hline \mathbf{0} & \mathbf{0} & \mathbf{M} - \mathbf{N} \end{array} \right) = k + \mathrm{rank}(\mathbf{M} - \mathbf{N})
$$

since the matrix $(\mathbf{1}|\mathbf{I}'_k)$ has rank $k$, and hence $d_{\mathrm{M}}(I_{\mathcal{A}}(\mathbf{M}), I_{\mathcal{A}}(\mathbf{N})) = d_{\mathrm{R}}(\mathbf{M}, \mathbf{N})$. ∎

We now design a class of nearly optimal codes for the affine geometry based on *affine liftings of Gabidulin codes*. Let $C$ be a Gabidulin code on $\mathrm{GF}(q)^{k \times (n-k+1)}$ with minimum rank distance $d$. Then by Proposition 11, its affine lifting $I_{\mathcal{A}}(C) = \{I_{\mathcal{A}}(\mathbf{M}) : \mathbf{M} \in C\}$ is a matroid code of $\mathcal{A}(q,n)$ with rank $k$, minimum distance $d$, and cardinality $q^{\min\{(n-k+1)(k-d+1),k(n-k-d+2)\}}$.

*Corollary 3:* For all $0 \leq k \leq n+1$, we have $A(\mathcal{A},k,d) \geq q^{\min\{(n-k+1)(k-d+1),k(n-k-d+2)\}}$.

As a corollary of this construction, we obtain $A(\mathcal{A},k,d) \geq q^{\min\{(n-k+1)(k-d+1),k(n-k-d+2)\}}$ for all $0 \leq k \leq n+1$. Combining this result with (14) and the bounds on the maximum cardinality of constant-dimension codes reviewed in Section II, we obtain $K_q < \frac{A(\mathcal{A}(q,n),k,d)}{A(\mathcal{L}(q,n+1),k,d)} \leq 1$. Therefore, RANC utilizes codes with a similar cardinality to subspace codes for packets longer by one symbol. This gain is clearly illustrated by the definition of affine lifting, which removes the first column from the identity matrix used in the linear lifting. The gain in data rate in (2) derived for the error-free case is hence preserved when error control is implemented. Furthermore, we prove below that this gain comes with no significant cost in terms of decoding complexity.

By construction, the lifting introduced above for RANC is closely related to the lifting introduced in [1] for RLNC. We now utilize this relation to design a low-complexity decoding of affine liftings of Gabidulin codes. More generally, we prove that decoding the affine lifting of a rank metric code can be performed using a subspace distance decoder for the linear lifting of the same code.

In order to clarify notations, we shall use the subscripts $\mathcal{A}$ and $\mathcal{L}$ to refer to objects (ranks and lattice distances) defined for RANC and RLNC, respectively. We introduce the nonsingular matrix $\mathbf{X}_{n+1} = (\mathbf{v}_k^T | \mathbf{I}'_{n+1}) \in \mathrm{GF}(q)^{(n+1)\times(n+1)}$, where $\mathbf{v}_k = (1, -1, -1, \ldots, -1, 0, \ldots, 0)$ has $k$ nonzero coefficients. For any affine subspace $M \in \mathcal{F}(\mathcal{A}(q, n))$ of rank $k$ given by the closure of the rows of the matrix $\mathbf{M} \in \mathrm{GF}(q)^{k\times n}$, we denote the linear subspace of $\mathrm{GF}(q)^{n+1}$ with dimension $k$ generated by $(\mathbf{1}|\mathbf{M})\mathbf{X}_{n+1}$ as $r(M) \in \mathcal{F}(\mathcal{L}(q, n+1))$. Multiplying on the right by $\mathbf{X}_{n+1}$ can hence be viewed as mapping the affine subspaces of $\mathrm{GF}(q)^n$ into linear subspaces of $\mathrm{GF}(q)^{n+1}$. Proposition 12 below shows that this mapping preserves the lattice distance, and that the image of the affine lifting of a matrix is the linear lifting of the same matrix.

*Proposition 12:* For any affine subspace $M \in \mathcal{F}(\mathcal{A}(q, n))$ and any affine lifting $I_{\mathcal{A}}(\mathbf{C}) \in \mathcal{F}(\mathcal{A}(q, n))$, we have $d_{\mathrm{L},\mathcal{A}}(M, I_{\mathcal{A}}(\mathbf{C})) = d_{\mathrm{L},\mathcal{L}}(r(M), I_{\mathcal{L}}(\mathbf{C}))$.

*Proof:* Since $\mathbf{X}_{n+1}$ is nonsingular, we have $\mathrm{rk}_{\mathcal{A}}(M) = \mathrm{rank}\{(\mathbf{1}|\mathbf{M})\mathbf{X}_{n+1}\} = \mathrm{rk}_{\mathcal{L}}(r(M))$. Also, it is easily shown that $(\mathbf{1}|\mathbf{I}'_k|\mathbf{C})\mathbf{X}_{n+1} = (\mathbf{I}_k|\mathbf{C})$, and hence

$$d_{\mathrm{L},\mathcal{A}}(M, I_{\mathcal{A}}(\mathbf{C})) = 2\mathrm{rank}\left(\begin{array}{c|c} \mathbf{1} & \mathbf{M} \\ \hline \mathbf{1} & \mathbf{I}'_k \mid \mathbf{C} \end{array}\right) - \mathrm{rank}(\mathbf{1}|\mathbf{M}) - \mathrm{rank}(\mathbf{1}|\mathbf{I}'_k|\mathbf{C}) \qquad (17)$$

$$= 2\mathrm{rank}\left(\frac{(\mathbf{1}|\mathbf{M})\mathbf{X}_{n+1}}{(\mathbf{I}_k|\mathbf{C})}\right) - \mathrm{rank}\{(\mathbf{1}|\mathbf{M})\mathbf{X}_{n+1}\} - \mathrm{rank}(\mathbf{I}_k|\mathbf{C}) \qquad (18)$$

$$= d_{\mathrm{L},\mathcal{L}}(r(M), I_{\mathcal{L}}(\mathbf{C})), \qquad (19)$$

where (17) and (19) follow the definition of the lattice distance in (7), while (18) is obtained by multiplying by $\mathbf{X}_{n+1}$ on the right. ∎

By Proposition 12, decoding $M$ using the affine lifting of a Gabidulin code is equivalent to decoding $r(M)$ using the linear lifting of the same code. We remark that transforming the matrix $\mathbf{M}$ into $(\mathbf{1}|\mathbf{M})\mathbf{X}_{n+1}$ can be simply performed by adding a column in front of $\mathbf{M}$, whose value is given by $\mathbf{1}-\sum_{i=0}^{k-2}\mathbf{col}_i$, where $\mathbf{col}_i$ denotes the $i$-th column of $\mathbf{M}$ for all $0 \le i \le n-1$. Therefore, the decoding algorithm for the affine lifting of a Gabidulin code follows two steps: First, obtain a matrix $\mathbf{M}$, and add the column $\mathbf{1}-\sum_{i=0}^{k-2}\mathbf{col}_i$ in front of it; Second, apply the bounded subspace distance decoding algorithm in [1] for the row space of the extended matrix. It is clear that the complexity of this algorithm is on the same order as that of the algorithm in [1] for the same Gabidulin code, which is $O(n^2)$ operations over $\mathrm{GF}(q)^{n-k+1}$. In order to summarize our results, the proposed implementation scheme for affine network coding is illustrated in Figure 4 for $2k \le n+1$.

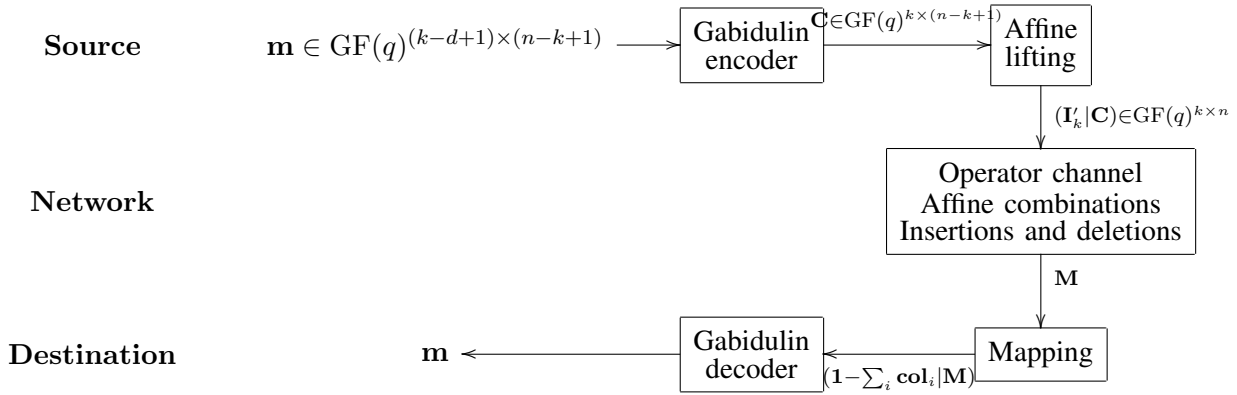

Fig. 4.   Implementation scheme for affine network coding

## VII. Conclusion

In this paper, we introduced a novel model for the performance study of and noncoherent error control for data transmissions through a network. This model, based on flats of matroids, encompasses traditional techniques, such as linear network coding and routing, and offers a wealth of alternatives to these protocols. We evaluated the performance of these two protocols both in the error-free case and in the case where packets are lost, injected, or in error. We then designed a new network coding protocol based on the affine geometry which outperforms linear network coding in terms of data rate for the coded and non-coded cases. We identified a class of nearly optimal codes, for which we provide a low-complexity decoding algorithm. The results are summarized in Table I.

This topic opens many directions for future research, some of which are detailed below. First, the model we proposed is based on simple assumptions, which may not accurately reflect the reality of the network.

| | Protocol | SAF | RLNC | RANC |
|---|---|---|---|---|
| Matroid Parameters | Matroid | $U_{q^n,q^n}$ | $PG(n-1,q)$ | $AG(n,q)$ |
| | Rank of the matroid | $q^n$ | $n$ | $n+1$ |
| | Combination | Selection | Linear combination | Affine combination |
| | Flats | Subsets | Linear subspaces | Affine subspaces |
| | Number of flats of rank $k$ | $\binom{q^n}{k}$ | $\begin{bmatrix} n \\ k \end{bmatrix}$ | $q^{n-k+1}\begin{bmatrix} n \\ k-1 \end{bmatrix}$ |
| | Cardinality of flats of rank $k$ | $k$ | $\frac{q^k-1}{q-1}$ | $q^{k-1}$ |
| Performance Parameters | Rate | $\sim 1-\frac{\log_q k}{n}$ | $\sim 1-\frac{k}{n}$ | $\sim 1-\frac{k-1}{n}$ |
| | Average delay | $\sim k\log k$ | $\sim k$ | $\sim k$ |
| | Throughput | $\sim \frac{1}{\log k}-\frac{1}{n\log q}$ | $\sim 1-\frac{k}{n}$ | $\sim 1-\frac{k-1}{n}$ |
| | Independent elements | $\sim k(1-e^{-\frac{r}{k}})$ | $\sim \min\{r,k\}$ | $\sim \min\{r,k\}$ |
| | Partially decodable elements | $l$ | $\sim kq^{l-k}$ | $kq^{l-k}$ |

TABLE I

Summary of parameters for SAF, RLNC, and RANC.

Hence, we need to investigate how the specificity of the given network can be incorporated into our model. Second, many different types of matroids have been proposed, from the most elementary to the most sophisticated. Determining which matroids are desirable for a given situation is an important research direction, as it also determines the corresponding protocol. Third, once the matroid is fixed, some tools are required to evaluate its performance and to compare it with other matroids. Although we introduced some parameters, such as the data rate and the average delay, new parameters may reflect some situations more accurately. Fourth, random affine network coding deserves to be investigated in further detail. In particular, the implementation of the low-complexity decoding procedure for liftings of Gabidulin codes introduced in this paper has a significant impact on the feasibility of affine network coding. Fifth, on a more practical approach, combining matroids may take advantage of the original matroids. For instance, combining SAF and RLNC may lead to transmitting packets with different priorities. in terms of error control, combining matroids may also lead to unequal protection against errors.

## VIII. Acknowledgment

The authors would like to acknowledge the Associate Editor Professor Frank R. Kschischang and the anonymous reviewers for their valuable comments.